\documentclass[aps,prl,twocolumn,letterpaper,superscriptaddress,showpacs]{revtex4-1}
\usepackage{keyval}%
\usepackage{graphicx}
\usepackage{dcolumn}
\usepackage{bm}
\usepackage{color}
\usepackage{amsmath,amssymb}
\usepackage{soul}

\begin{document}

\title{Peculiar bonding associated with atomic doping and hidden honeycombs in borophene}
\author{Chi-Cheng Lee}
\affiliation{Institute for Solid State Physics, The University of Tokyo, Kashiwa, Chiba 277-8581, Japan}%
\author{Baojie Feng}
\affiliation{Institute for Solid State Physics, The University of Tokyo, Kashiwa, Chiba 277-8581, Japan}%
\author{Marie D'angelo}
\affiliation{Institute for Solid State Physics, The University of Tokyo, Kashiwa, Chiba 277-8581, Japan}%
\affiliation{Institut des Nanosciences de Paris, Universit\'{e} Pierre et Marie Curie, CNRS 4, place Jussieu, 75005 Paris, France}%
\author{Ryu Yukawa}
\affiliation{Institute of Materials Structure Science, High Energy Accelerator Research Organization (KEK), Tsukuba, Ibaraki 305-0801, Japan}%
\author{Ro-Ya Liu}
\affiliation{Institute for Solid State Physics, The University of Tokyo, Kashiwa, Chiba 277-8581, Japan}%
\author{Takahiro Kondo}
\affiliation{Tsukuba Research Center for Energy Materials Science (TREMS), University of Tsukuba, Tsukuba, 305-8571, Japan}
\affiliation{Division of Materials Science, Faculty of Pure and Applied Sciences, University of Tsukuba, Tsukuba 305-8573, Japan}
\affiliation{Materials Research Center for Element Strategy, Tokyo Institute of Technology, Yokohama 226-8503, Japan}
\author{Hiroshi Kumigashira}
\affiliation{Institute of Materials Structure Science, High Energy Accelerator Research Organization (KEK), Tsukuba, Ibaraki 305-0801, Japan}
\author{Iwao Matsuda}
\affiliation{Institute for Solid State Physics, The University of Tokyo, Kashiwa, Chiba 277-8581, Japan}%
\author{Taisuke Ozaki}
\affiliation{Institute for Solid State Physics, The University of Tokyo, Kashiwa, Chiba 277-8581, Japan}%
\date{\today}

\begin{abstract}
Engineering atomic-scale structures allows great manipulation of physical properties and chemical processes for advanced technology. 
We show that the B atoms deployed at the centers of honeycombs in boron sheets, borophene, behave as nearly perfect electron donors
for filling the graphitic $\sigma$ bonding states without forming additional in-plane bonds by first-principles calculations. The dilute electron density distribution
owing to the weak bonding surrounding the center atoms provides easier atomic-scale engineering and is highly tunable via in-plane strain, promising for practical 
applications, such as modulating the extraordinarily high thermal conductance that exceeds the reported value in graphene. The hidden honeycomb bonding structure suggests 
an unusual energy sequence of core electrons that has been verified by our high-resolution core-level photoelectron spectroscopy measurements. With the experimental and 
theoretical evidence, we demonstrate that borophene exhibits a peculiar bonding structure and is distinctive among two-dimensional materials.
\end{abstract}
  
\maketitle

Graphene, the representative of two-dimensional materials, has been proposed for various applications, such as nanoelectronics and optoelectronics 
for the next generation of technology\cite{Ref1,Ref2,Ref3}. Not only hosting massless Dirac fermions makes it attractive but also the robust bonding giving a 
remarkable stiffness renders the honeycomb structure one of the most attractive patterns in materials science\cite{Ref1,Ref2,Ref3}. Similar to carbon, boron has been found 
to exist in a variety of structures associated with multicenter bonding, where the bonds involve multiple atoms sharing a certain amount of electrons, 
ranging from clusters to bulks\cite{Ref4,Ref5,Ref6}. The flexible bonding nature provides the degrees of freedom of atomic-scale engineering for great manipulation of 
physical properties and chemical processes, especially in the layer forms that can be grown on diverse substrates. The boron layer exhibits many interesting properties. 
For example, the graphitic boron layer in MgB$_2$ has set a remarkable record for the superconductivity transition temperature ($T_c\sim$40K) among simple binary 
compounds\cite{Ref7}, making the two-dimensional boron layer ($T_c\sim$20K) lastingly attractive for realizing better superconductors\cite{Ref8,Ref9}. 
Exploration of new boron compounds to keep pace with the graphene technology has also been ongoing\cite{Ref10}. Recently two-dimensional boron sheets, borophene, 
have attracted great attention due to the successful growth on a metallic substrate\cite{Ref11,Ref33,Ref12,Ref13,Ref14,Ref15,Ref16,Ref17,Ref18,Ref19,Ref20,Ref21}. 
Dirac cones were also evidenced in borophene\cite{Ref22,Ref23,Ref24}. These make borophene another promising candidate for manufacturing advanced nanoscale devices. 
It is then interesting to unravel the bonding nature of borophene, which is composed of mixtures of honeycombs and triangles, and to propose useful applications 
with physical properties superior to graphene. 

\begin{figure}[tbp]
\includegraphics[width=1.00\columnwidth,clip=true,angle=0]{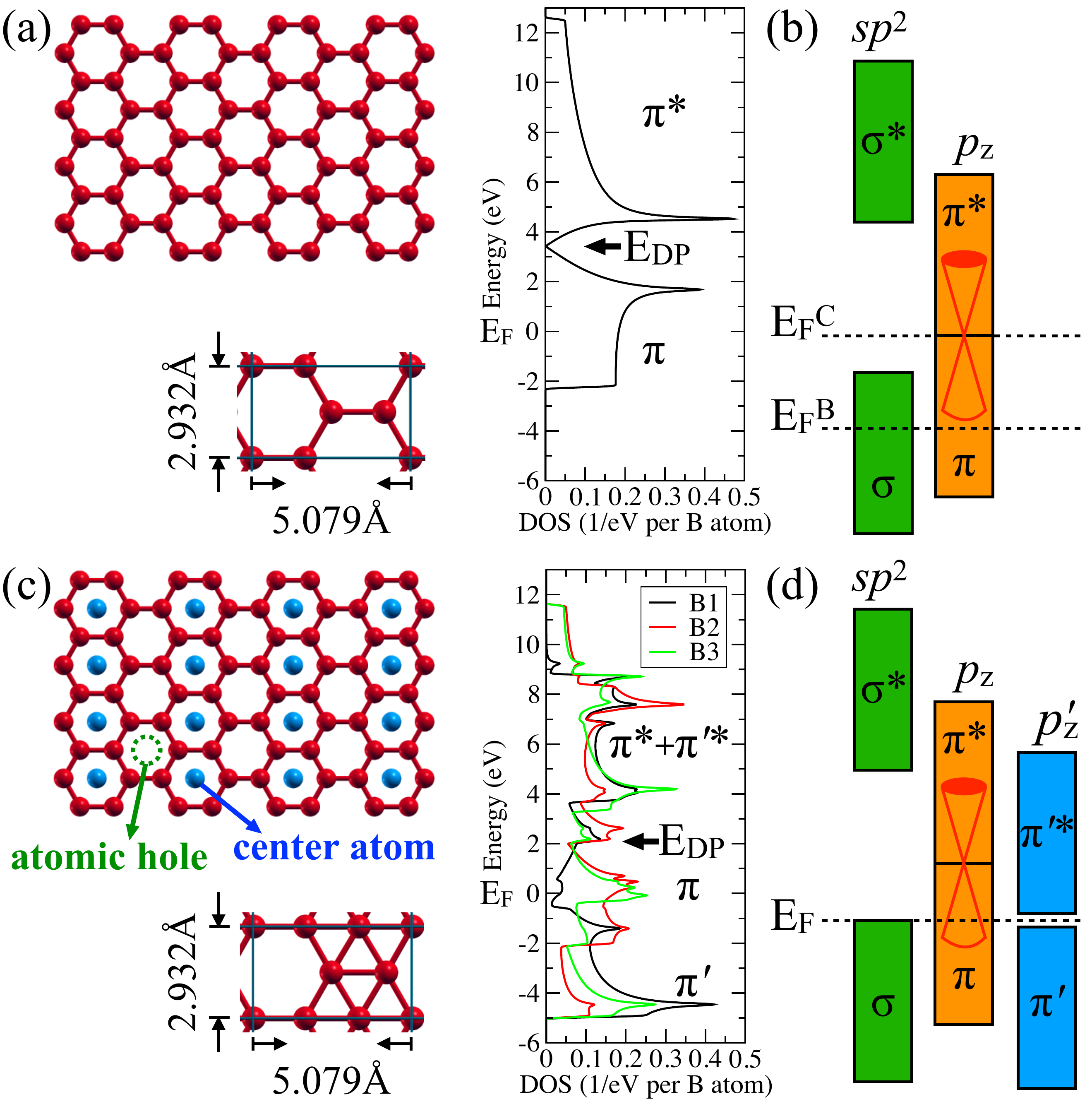}
\caption{
Geometrical structures and density of states of B $p_z$ orbitals (DOS) obtained from first-principles calculations on freestanding planar (a) honeycomb
and (c) $\beta_{12}$ sheets of borophene. The energies of Dirac points ($E_{DP}$) are indicated by arrows.
(b) Sketches of filling of $sp^2$ and $p_z$ orbitals in the carbon and boron honeycombs with the Fermi levels at $E_F^C$ and $E_F^B$, respectively,
together with (d) a new arrangement of band filling with additional B atoms at honeycomb centers.
}
\label{fig:energylevel}
\end{figure}

The structures of borophene can be considered as introducing either vacancies, dubbed as atomic holes, or buckling to the prototypical planar 
triangular structure\cite{Ref11,Ref33,Ref12}. Alternatively, the atomic-hole structures can be viewed as adding and/or removing B atoms based on 
the graphitic honeycombs\cite{Ref11,Ref33}. The intrinsic difference between the boron and carbon versions of honeycombs is that four valence electrons per carbon atom 
optimally fill the bonding $\sigma$ and $\pi$ bands with exactly empty antibonding $\sigma^*$ and $\pi^*$ bands separated by a gap and Dirac points at the Fermi level, 
respectively, in graphene, whereas boron with one less electron cannot fully fill 
all bonding states\cite{Ref11,Ref33,Ref12,Ref13,Ref14,Ref15,Ref16,Ref17,Ref18,Ref19,Ref20,Ref21,Ref22,Ref23,Ref24}. Hence, the density of atomic holes 
is intimately associated with an electron-doping mechanism between the two-center and three-center bonding for stabilizing borophene by noting that the three-center bonding 
in the triangular structure possesses excess electrons, which has been demonstrated by the first-principles calculations\cite{Ref11,Ref33}. 
The structures, density of states, and schematic pictures of band filling of honeycomb and $\beta_{12}$ sheets of borophene are shown in Fig.~\ref{fig:energylevel}.
The $\beta_{12}$ borophene that has been experimentally realized and theoretically explored very recently allows us to verify the doping mechanism and bonding nature 
predicted for the boron sheets in general\cite{Ref11,Ref33}. 

In this Communication, we focus on $\beta_{12}$ borophene that contains honeycombs plus additional B atoms at the honeycomb centers and has been experimentally realized 
on Ag(111)\cite{Ref23}. The relaxed structure is shown in Fig.~\ref{fig:energylevel} (c), where the center atoms are present in one column along the 
zigzag but absent in the next column. We will demonstrate that there exist novel effects associated with atomic doping 
by our first-principles calculations. In particular, we will show the center atoms behave as nearly 
perfect electron donors in filling the honeycomb $\sigma$ states without forming new in-plane bonds, allowing easier engineering of atomic-scale structures 
and great tunability of the surrounding charge density distribution. A peculiar $\pi$ bond shared by the center atom and the six atoms forming the honeycomb is found, 
which is beyond the picture of mixed two-center and three-center bonding and can be considered as six-center bonding by viewing the center atom as a pure electron
reservoir. Finally, we will provide experimental evidence of the hidden honeycomb bonding structure in borophene. 
The experimental and computational details can be found elsewhere\cite{Supp}.

\begin{figure}[tbp]
\includegraphics[width=1.00\columnwidth,clip=true,angle=0]{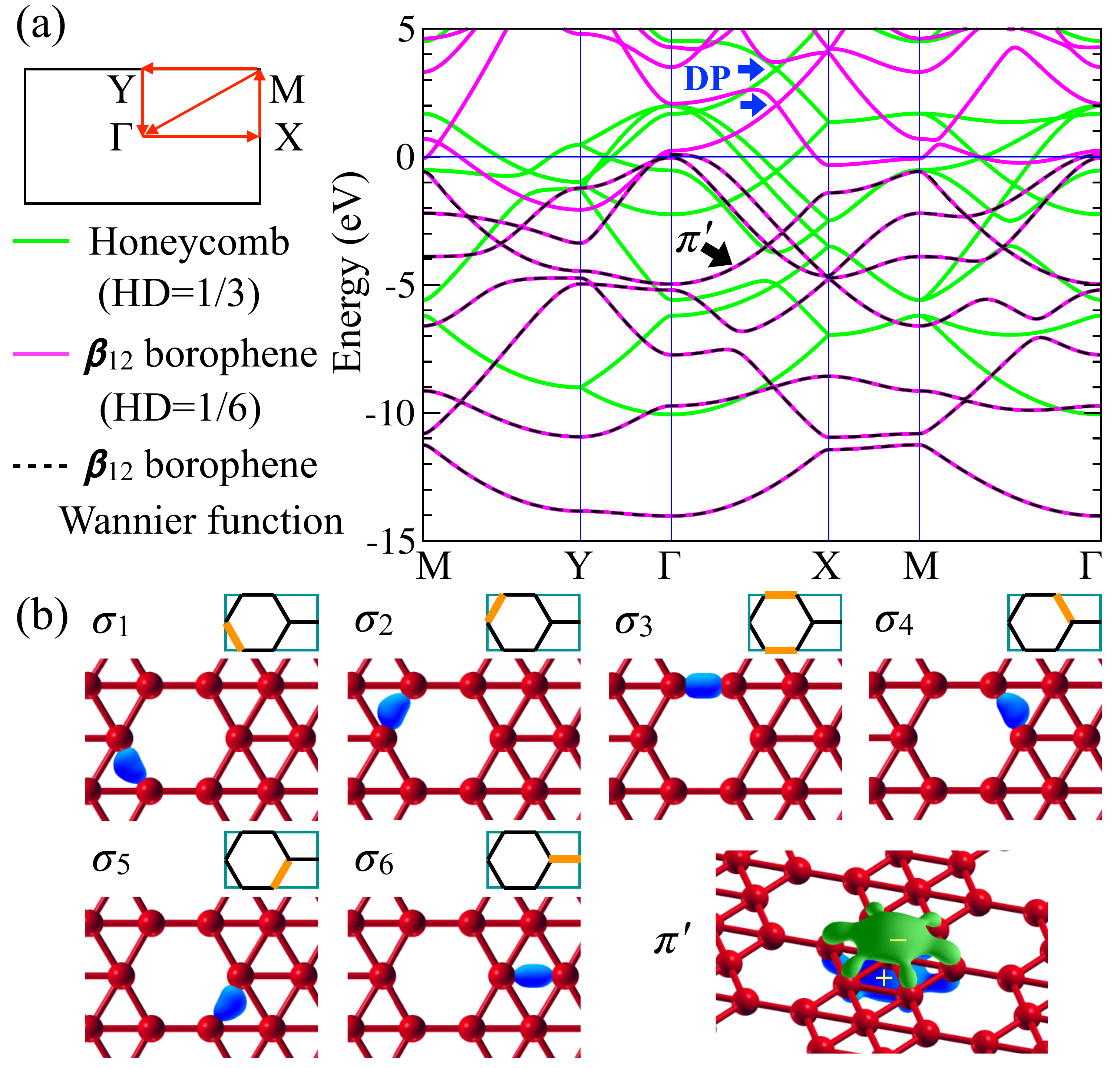}
\caption{
(a) First-principles band structures of honeycomb and $\beta_{12}$ sheets of borophene at the atomic hole density (HD) of 1/3 and 1/6, 
respectively. The Dirac points (DP) are indicated by arrows. The dashed curves are generated from the tight-binding Hamiltonian in the basis 
of (b) the Wannier functions of $\beta_{12}$ borophene shown with the isosurfaces at 0.23 $\sqrt{e/Bohr^3}$ for the $\sigma$ orbitals and 0.07 $\sqrt{e/Bohr^3}$ 
for the $\pi^\prime$ orbital. The band dispersion of $\pi^\prime$ orbital is indicated in (a).
}
\label{fig:bandstructure}
\end{figure}

The first-principles band structures of fully relaxed $\beta_{12}$ borophene and the corresponding honeycomb version are shown in Fig.~\ref{fig:bandstructure} (a). 
The relaxed lattice constant of honeycomb sheet is just $\sim$0.4\% shorter than that of $\beta_{12}$ sheet so that the presence of center atoms does not modify the 
honeycomb size significantly. As expected, the $\sigma$ bands are not fully filled in the honeycomb structure, as evidenced by the downward bands right above the Fermi level 
at $\Gamma$. On the other hand, the $\sigma$ bands become nearly fully filled in $\beta_{12}$ borophene as an evidence of electron doping via the center B atoms. 
The crossing right above the Fermi level at the point located at the 2/3 of $\Gamma$ to $X$ path corresponds to the Dirac point at $K$ in the primitive Brillouin zone of 
honeycomb borophene. Such a Dirac cone also exists in $\beta_{12}$ borophene and the Dirac fermions can be observed in angle-resolved photoelectron spectroscopy experiments
by further electron doping\cite{Ref23}. 

\begin{figure*}[tbp]
\includegraphics[width=2.00\columnwidth,clip=true,angle=0]{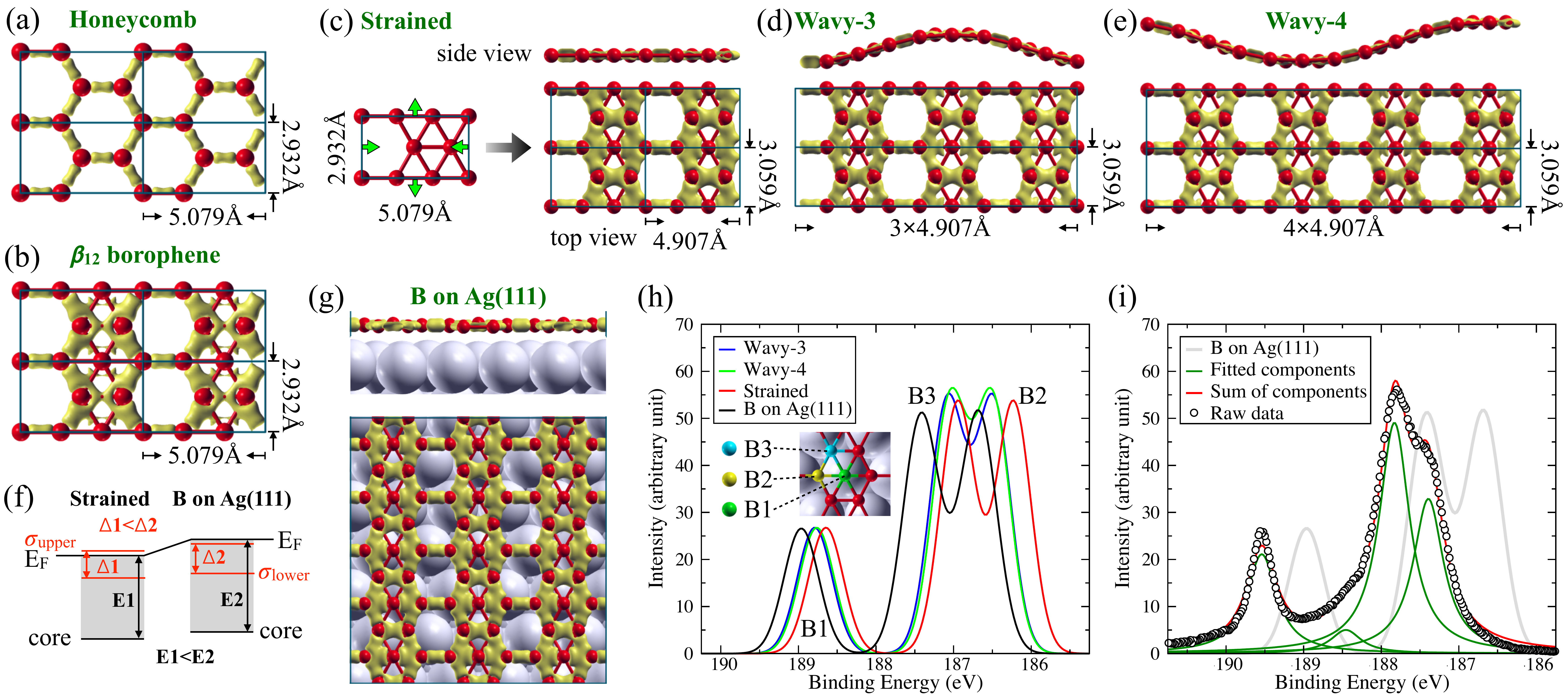}
\caption{
Isosurfaces of charge density at 0.125 $e/Bohr^3$ in planar (a) honeycomb and (b) relaxed $\beta_{12}$ borophene.
Isosurfaces of charge density at 0.12 $e/Bohr^3$ in (c) planar, (d) shorter-wavelength wavy, and (e) longer-wavelength wavy borophene in the strained 
in-plane unit cells commensurate with $\beta_{12}$ borophene on Ag(111) whose isosurface is shown in (g). (f) The energies of Fermi levels ($E_F$), core levels (core), 
and upper bound ($\sigma_{upper}$) and lower bound ($\sigma_{lower}$) of $\sigma$ bands at $\Gamma$ of freestanding strained $\beta_{12}$ borophene and 
$\beta_{12}$ borophene on Ag(111) are sketched. (h) First-principles binding energies of B 1$s$ orbitals at different sites weighted by the associated number of atoms 
in the unit cells. (i) Measured high-resolution core-level photoelectron spectroscopy spectra of $\beta_{12}$ borophene on Ag(111) together with the fitted components 
on top of the theoretical result.
}
\label{fig:xps}
\end{figure*}

To unravel the bonding nature in $\beta_{12}$ borophene, the maximally localized Wannier functions\cite{Ref25,Ref26} transformed from the seven dominant occupied bands and 
the reproduced bands are presented in Figs.~\ref{fig:bandstructure} (b) and (a), respectively. Six $\sigma$ orbitals that are translationally invariant can be seen 
forming the honeycombs, revealing the hidden honeycomb bonding structure that has also been found in other boron sheets\cite{Ref33}. 
Another evidence is the similar electron density 
distribution between honeycomb and $\beta_{12}$ sheets as shown in Figs.~\ref{fig:xps} (a) and (b), respectively. Only dilute charge density can be found around the center atom 
with a larger area of the honeycomb isosurface of charge density in $\beta_{12}$ borophene, reflecting that the center atom behaves as a nearly perfect electron donor for filling 
the honeycomb $\sigma$ bonds. 

The remaining Wannier function can be identified as the $p_z$ orbital of the dopant atom hybridizing with neighboring $p_z$ orbitals that can be considered as 
six-center (or seven-center by taking the dopant atom into account) bonding filled by two electrons in the space orthogonal to the $\sigma$ bonds, corresponding to the $\pi^\prime$ band with a gap to the other $p_z$-derived bands
as shown in Fig.~\ref{fig:bandstructure} (a).  
The other partially filled bands that are not represented by the Wannier functions originate from the $p_z$ orbitals of honeycomb B atoms.
Specifically, two $\pi$ and two $\pi^*$ bands can be obtained in the honeycomb borophene by doubling the unit cell. Adding one center atom 
that breaks the translational symmetry of honeycomb structure gives five $p_z$ bands from the five mutually hybridized $p_z$ orbitals. As shown in Fig.~\ref{fig:energylevel},
the degeneracy in the energy distribution of the original $\pi$ bands is lifted and a new lower-energy $\pi$ contribution, the $\pi^\prime$ band, can be identified. 
The additional band is found to hybridize more with the original two $\pi^*$ bands, leading to three 
nearly fully unfilled $\pi^*+\pi^{\prime*}$ bands in $\beta_{12}$ borophene. 

The dopant atoms are solely bonded by the $\pi^\prime$ orbitals. Besides the weak $\pi^\prime$ bonding, 
dilute in-plane charge density can still spread surrounding the center atoms reflected by the deformed $\sigma_1$, $\sigma_2$, $\sigma_4$, and $\sigma_5$ orbitals in 
comparison to the $\sigma_3$ and $\sigma_6$ orbitals having no tails approaching the honeycomb center in Fig.~\ref{fig:bandstructure} (b). The weak $\pi^\prime$ bond and the dilute in-plane 
charge density imply a highly tunable electron density distribution surrounding the center atom via strain, which could directly affect the properties of the electronic structure and 
phonons. In Fig.~\ref{fig:xps} (c), we show the electron density distribution in a strained unit cell,  
where the corresponding (3$\times$5) unit cell can fit (5$\times$6) Ag(111) in the rectangular supercell. As expected, the electron is distributed more along the 
shorter bonds and less along the longer bonds measured from the center atom, building a new channel having an interesting one-dimensional electron density distribution along 
the shorter-bond direction under non-uniform strain.

Substrate-induced undulations in $\beta_{12}$ borophene have been observed on Ag(111) with existence of additional protruding Ag atoms\cite{Ref13}. Here we show that strain 
can also induce undulations without the presence of the substrate by focusing on sinusoidal sheets at two different wavelengths. While the two wavy structures have similar total energies, 
their total energies are lower than that of the strained planar sheet by the order of 10 meV per atom as an energy gain from relaxing the imposed in-plane stress. 
As shown in Figs.~\ref{fig:xps} (d) and (e), the feature of one-dimensional electron density distribution can also be found. This is useful for practical applications because charge density 
around the center atom can be controlled by the in-plane strain and is robust against undulations. With the presence of silver, the interfacial cohesive energy of borophene on Ag(111)\cite{Ref17}, 
$\sim$0.17 eV per B atom, is larger than the energy gain from the sinusoidal forms. As a result, a nearly planar 
sheet can be found as shown in Fig.~\ref{fig:xps} (g), where the feature of strain-induced one-dimensional electron density distribution is again observed. 
In addition, prominent honeycomb electron density distribution is always observed in all the cases, showing the robustness of the honeycomb bonding structure 
against structural flexibility.

The peculiar honeycomb bonding implies an unusual energy sequence of core electrons that can be verified by high-resolution photoelectron spectroscopy experiments. 
The coordination number of center B atoms is six, where much stronger Coulomb repulsion and therefore shallower site energy of B 1$s$ orbitals to 
the Fermi level can be expected. In the independent-electron picture, 
the core-level binding energy is the energy difference between the site energy and the Fermi level. So the binding energy of B 1$s$ orbital at the center atom should be 
the smallest among all the B atoms. However, the unexpectedly dilute charge density has been found to surround the center atom in the graphitic honeycomb bonding. Consequently, 
the binding energy belonging to the center atoms should be the largest instead of the smallest. 

To confirm the unusual energy sequence, the calculated absolute binding energies of three distinct B 1$s$ orbitals, denoted as B1, B2, and B3 shown in Fig.~\ref{fig:xps} h, 
are listed in Table~\ref{table:bindingenergy}. In all cases, including the relaxed planar sheet, strained planar and undulation sheets, and the nearly planar sheet on Ag(111), 
the B1 1$s$ binding energies are prominently larger than those at the other sites. Comparing to the planar sheet, the undulations give larger binding energy 
for each respective B atom as a result of longer bond lengths reducing both Coulomb repulsion and in-plane strain. The similar electron density distribution of $\beta_{12}$ borophene on Ag(111) 
preserves the same energy sequence as in the freestanding cases. Due to the charge transfer from Ag(111) to borophene and the interaction between them\cite{Ref16,Ref23}, 
the relatively higher Fermi level in the presence of silver gives larger binding energy as illustrated in Fig.~\ref{fig:xps} (f). 
The calculated single-particle energy of the lowest B1 1$s$ level and the Fermi level before being shifted to the energy zero in the strained borophene are -6.621 and -0.202 Ha, 
and become -6.600 and -0.165 Ha with the presence of silver, respectively, where $\sim$0.4 eV binding energy is increased. 
The energy sequence can be further understood by counting the number of bonds 
surrounding the core electrons following the electron density distribution instead of the coordination number since the number is approximately proportional to the strength of Coulomb repulsion. 
As shown in Figs.~\ref{fig:xps} (c), (d), (e), and (g), the numbers of B1, B2, and B3 are two, four, and three, respectively, perfectly matching the energy sequence 
listed in Table~\ref{table:bindingenergy}. 

\begin{table}[tbp]
\caption{
Binding energies of B 1s orbitals in relaxed planar $\beta_{12}$ borophene (Relaxed), planar (Strained) and wavy ones in a strained unit cell, 
and the relaxed one on Ag(111). The wavelengths of three (Wavy 3) and four (Wavy 4) times the longer lattice constant are considered. The average values 
are listed for broken symmetry cases together with the fitted components (Exp)\cite{Supp}. The unit is in eV.
}
\label{table:bindingenergy}%
\begin{tabular}{ccccccc}
\hline
         & Relaxed  & Strained & Wavy 3 & Wavy 4 & Ag(111) & Exp  \\
\hline
B1       & 188.567 & 188.646 & 188.789 & 188.765 & 188.962 & 189.538 \\
B2       & 186.331 & 186.228 & 186.488 & 186.496 & 186.678 & 187.391 \\
B3       & 186.757 & 186.946 & 187.081 & 187.050 & 187.412 & 187.828 \\ \hline
\end{tabular}%
\end{table}

The experimentally measured B 1$s$ binding energies that support the peculiar honeycomb bonding in $\beta_{12}$ borophene on Ag(111) are presented in Table~\ref{table:bindingenergy} and Fig.~\ref{fig:xps} (i), 
where the prominent higher-energy B1 peak and lower-energy B2 and B3 peaks can be clearly observed. To fit the measured raw data, at least one additional small peak is required. 
Although $\sim$0.5 eV deviation could be obtained in the first-principles calculations of absolute binding energies\cite{Ref28,Ref29}, it is possible that additional degrees of freedom 
not considered in the supercell calculations, such as defects, randomly distributed center atoms, undulations, domain boundaries\cite{Ref15,Ref16}, and other strain-relaxed forms could give better agreement 
between theory and experiment. While such exploration is interesting, the universal energy sequence of the core electrons cannot be easily altered due to the demonstrated 
robustness of the honeycomb bonding structure, and the three major peaks in the raw data should mainly come from B1, B2, and B3 of the extended $\beta_{12}$ borophene. 

Finally, we mention that the buckled triangular and $\beta_{12}$ sheets of borophene have extraordinarily high lattice thermal conductance exceeding that of graphene\cite{Ref20}. 
In buckled borophene, the electron density distribution along the perfect one-dimensional chain is found to be responsible for the high-frequency phonon-mediated thermal transport\cite{Ref20}. 
The tunable dilute electron density distribution that we have evidenced for $\beta_{12}$ borophene may allow modulating the low-frequency phonons for highly tunable 
anisotropic thermal conductance via in-plane strain. Moreover, the one-dimensional electron density distribution demonstrated in Fig.~\ref{fig:xps} resembles that 
of buckled borophene, implying that an even higher thermal conductance can be realized under strain, which can be switched off by opposite strain. More applications associated 
with this flexible directional bonding, such as enhancing the electron-phonon coupling for better superconductors, are also expected.

In conclusion, we have identified a peculiar bonding structure in $\beta_{12}$ borophene.
The center B atom acts as a nearly perfect electron donor to fill the honeycomb $\sigma$ bonds in $\beta_{12}$ borophene. The newly introduced bond to the 
honeycomb structure is just a weak $\pi$-type six-center bond without additional stronger in-plane $\sigma$ bonds, which greatly facilitates atomic-scale engineering associated 
with the center atoms. The associated unusual core-level binding energy sequence owing to the unexpectedly dilute charge density 
surrounding the center atom has been verified by both first-principles calculations and high-resolution core-level photoelectron spectroscopy measurements. 
The weak $\pi$-type bonding and dilute in-plane charge density surrounding the center atom allow a highly tunable electron density distribution. 
A new channel having one-dimensional electron density distribution under in-plane strain is found and robust against undulations and the presence of a metallic substrate, 
useful for practical applications, such as modulating the anisotropic high thermal conductivity. More electron density distribution could be realized with different deployments of 
atomic holes, showing a playground for engineering and designing advanced devices. 

\begin{acknowledgments}
This work was supported by Priority Issue (creation of new functional devices and high-performance materials to support next-generation industries) to be tackled by using Post `K' Computer, Ministry of Education, Culture, Sports, Science and Technology, Japan.
\end{acknowledgments}

\bibliography{refs}

\end{document}